\documentclass[12pt]{article}
\usepackage{amsmath}
\usepackage{amssymb}
\usepackage{slashed}
\usepackage{graphicx}
\usepackage{subfigure}
\usepackage{cite}
\usepackage{url}
\usepackage[small]{caption}
\usepackage{multirow}
\usepackage{latexsym}
\usepackage{hyperref}
\usepackage{breakurl}

\hypersetup{colorlinks,citecolor=blue,linkcolor=black,urlcolor=black}

\begin{document}

\begin{titlepage}

%\begin{flushright}
%UCB-PTH-15/?? \\
%\end{flushright}

\hfill
\vspace{1.2cm}

\begin{center}
{\Large \bf 750 GeV Diphotons:\\
\vspace{1mm}
Implications for Supersymmetric Unification}

\vspace{1cm}

{\large Lawrence J. Hall, Keisuke Harigaya, and Yasunori Nomura}

\vspace{0.5cm}

{\it Berkeley Center for Theoretical Physics, Department of Physics, \\
 and Theoretical Physics Group, Lawrence Berkeley National Laboratory,\\
 University of California, Berkeley, CA 94720, USA}

\end{center}
\bigskip

\centerline{\large\bf Abstract}
\begin{quote}
A recent signal of $750~{\rm GeV}$ diphotons at the LHC can be explained 
within the framework of supersymmetric unification by the introduction 
of vector quarks and leptons with Yukawa couplings to a singlet $S$ 
that describes the $750~{\rm GeV}$ resonance.  We study the most 
general set of theories that allow successful gauge coupling unification, 
and find that these Yukawa couplings are severely constrained by 
renormalization group behavior: they are independent of ultraviolet 
physics and flow to values at the TeV scale that we calculate precisely. 
As a consequence the vector quarks and leptons must be light; typically 
in the region of $375~{\rm GeV}$ to $700~{\rm GeV}$, and in certain 
cases up to $1~{\rm TeV}$.  The $750~{\rm GeV}$ resonance may have 
a width less than the experimental resolution; alternatively, with 
the mass splitting between scalar and pseudoscalar components of $S$ 
arising from one-loop diagrams involving vector fermions, we compute 
an apparent width of 10s of GeV.
\end{quote}

\end{titlepage}

{\it If}\ the recently announced data on a $750~{\rm GeV}$ resonance 
decaying to diphotons~\cite{ATLAS,CMS} is confirmed, what are the 
implications for the framework of supersymmetric unification?  Neither 
the Minimal Supersymmetric Standard Model (MSSM) nor its extension to 
include a gauge singlet chiral multiplet can account for the resonance 
(see, however,~\cite{Allanach:2015ixl}).  In this paper we show that 
a simple addition to the theory, that maintains supersymmetric gauge 
coupling unification at $M_{\rm G} \sim 2 \times 10^{16}~{\rm GeV}$, 
can account for the resonance and leads to new vector-like quarks, 
with masses below $1~{\rm TeV}$ in the vast majority of parameter 
space, and couplings predicted from the infrared behavior of 
renormalization group equations.  (For the first studies of the 
$750~{\rm GeV}$ excess that appeared after the announcement, see 
Ref.~\cite{Harigaya:2015ezk}.)

We take the resonance to be the scalar component of a gauge singlet 
chiral multiplet $S$.  Production and decay of $S$ is accomplished by 
coupling to TeV-scale multiplets $\Phi_i$ and $\bar{\Phi}_i$, so that 
the effective theory below $M_{\rm G}$ is described by
\begin{equation}
  W_{\rm eff} =  W_{\rm MSSM} + S \sum_i \lambda_i\, \Phi_i \bar{\Phi}_i 
    + \frac{\mu_S}{2} S^2 + \mu_i\, \Phi_i \bar{\Phi}_i.
\label{eq:W}
\end{equation}
To preserve precision supersymmetric gauge coupling unification, 
we study theories where $\Phi_i$ and $\bar{\Phi}_i$ form complete 
multiplets of SU(5).  We study the complete set of such theories:\ 
the ``$({\bf 5} + \overline{\bf 5})_{N_5}$'' theory containing 
$N_5 = 1,2,3$ or $4$ copies of $(\bar{D},\bar{L}) + (D,L)$, the 
``${\bf 10} + \overline{\bf 10}$'' theory containing $(Q,U,E) + 
(\bar{Q},\bar{U},\bar{E})$, and the ``${\bf 15} + \overline{\bf 15}$'' 
theory that contains a full generation of vector quarks and leptons. 
In the $({\bf 5} + \overline{\bf 5})_4$ and ${\bf 15} + \overline{\bf 15}$ 
theories, the standard model gauge couplings near $M_{\rm G}$ are 
in the strong coupling regime, so they correspond to the scenario 
of strong ``unification''~\cite{Maiani:1977cg}, rather than precision 
perturbative unification, which applies to the other theories.

The parameters $\mu_S$ and $\mu_i$ are assumed to be of order the TeV 
scale, with an origin that may be similar to that of the supersymmetric 
Higgs mass parameter.  Any $S^3$ coupling in the superpotential is assumed 
to be sufficiently small not to affect our analysis.  The effective 
theory of Eq.~(\ref{eq:W}) possesses a parity on the vector matter, 
so that the lightest $\Phi_i$ is stable.  We therefore add Yukawa 
interactions between these vector quarks and leptons and the standard 
model quarks and leptons via the MSSM Higgs doublets; this can be done 
without violating $R$-parity if $\Phi_i$ and $\bar{\Phi}_i$ are $R$-parity 
odd.  We take these couplings to be sufficiently small that they do not 
affect the production and decay of the $750~{\rm GeV}$ resonance, and do 
not violate bounds on flavor-changing processes.  This allows the vector 
quarks and leptons to have prompt decays to known quarks and leptons 
with the emission of $W,Z$ or $h$, the $125~{\rm GeV}$ Higgs boson. 
(An alternative possibility will be discussed at the end of the paper.)

Soft supersymmetry breaking masses for the scalar components of $S, \Phi_i, 
\bar{\Phi}_i$ are present, as well as $A$ and $B$ type soft trilinears 
and bilinears.  For simplicity, these parameters are chosen so that $S$ 
does not acquire a vacuum expectation value.  Furthermore, we assume 
that the scalar components of $\Phi_i$ and $\bar{\Phi}_i$ are sufficiently 
heavy that they do not contribute significantly to the production 
or decay of $S(750)$.%
\footnote{For example, for the soft supersymmetry breaking masses of a TeV, 
 the contributions from the scalar components change the lower bounds on 
 $\mu_i$ by at most $10\%$.}
The bilinear scalar interaction, ${\cal L} \supset -b_S S^2/2 + 
{\rm h.c.}$, leads to a mass splitting between the two mass eigenstate 
scalars $S_1$ and $S_2$, leading to three distinct descriptions of the 
diphoton excess.  For a large mass splitting, the $750~{\rm GeV}$ state 
is described by $S_1$ alone, which has a narrow width much smaller 
than the experimental resolution.  If $b_S$ is one-loop suppressed 
the $S_1$-$S_2$ mass difference could be 10s of GeV, leading to an 
apparent width of this order~\cite{Franceschini:2015kwy}, which is 
mildly preferred by data.  Finally, smaller values for $b_S$ give 
a width below the experimental resolution.

The effective theory below $M_{\rm G}$ described by Eq.~(\ref{eq:W}) 
can result from a wide variety of unified theories.  Some unified 
theories have specific boundary conditions on $\lambda_i$ and $\mu_i$ 
at $M_{\rm G}$, for example $\lambda_D = \lambda_L$ and $\mu_D = \mu_L$ 
in the ``$({\bf 5} + \overline{\bf 5})_1$'' theory, with corrections of 
order $M_{\rm G}/M_*$ from higher dimension operators, where $M_*$ is 
the cutoff of the unified theory.  However, such boundary conditions 
could be absent, for example if $D$ and $L$ arise from different unified 
multiplets, as occurs in theories where boundary conditions in extra 
dimensions breaks the unified symmetry.  Below we make a simplifying 
assumption that the phases of $\lambda_i$ are common in the basis that 
$\mu_i$ are real and positive.  (This requires only coincidences of the 
signs if the relevant terms respect $CP$.)  Under this assumption, we 
find that the production and decay of $S(750)$ is insensitive to any 
boundary condition on $\lambda_i$, but critically dependent on boundary 
conditions for $\mu_i$.  Allowing arbitrary phases only reduces the 
diphoton signal, because of a possible cancellation in the amplitude 
for decays to diphotons, so that the upper bounds we derive on vector 
quark masses are conservative.  We stress that our analysis depends only 
on the effective theory, and does not depend on the unified gauge group or 
whether unification occurs in 4~dimensions~\cite{Dimopoulos:1981zb}, higher 
dimensions~\cite{Hall:2001pg}, or in string theory~\cite{Candelas:1985en}.

\begin{figure}[t]
\begin{center}
  \subfigure{\includegraphics[clip,width=.48\textwidth]{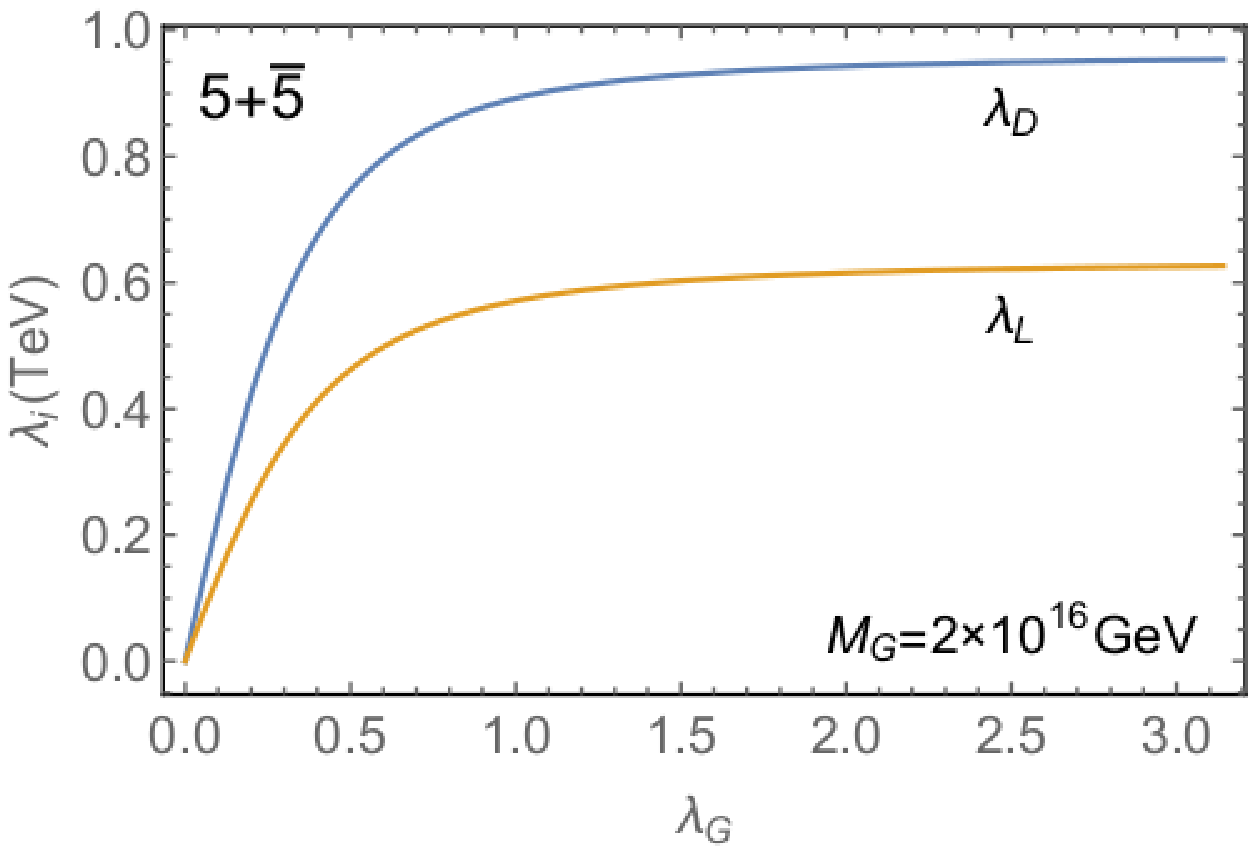}}
  \quad
  \subfigure{\includegraphics[clip,width=.48\textwidth]{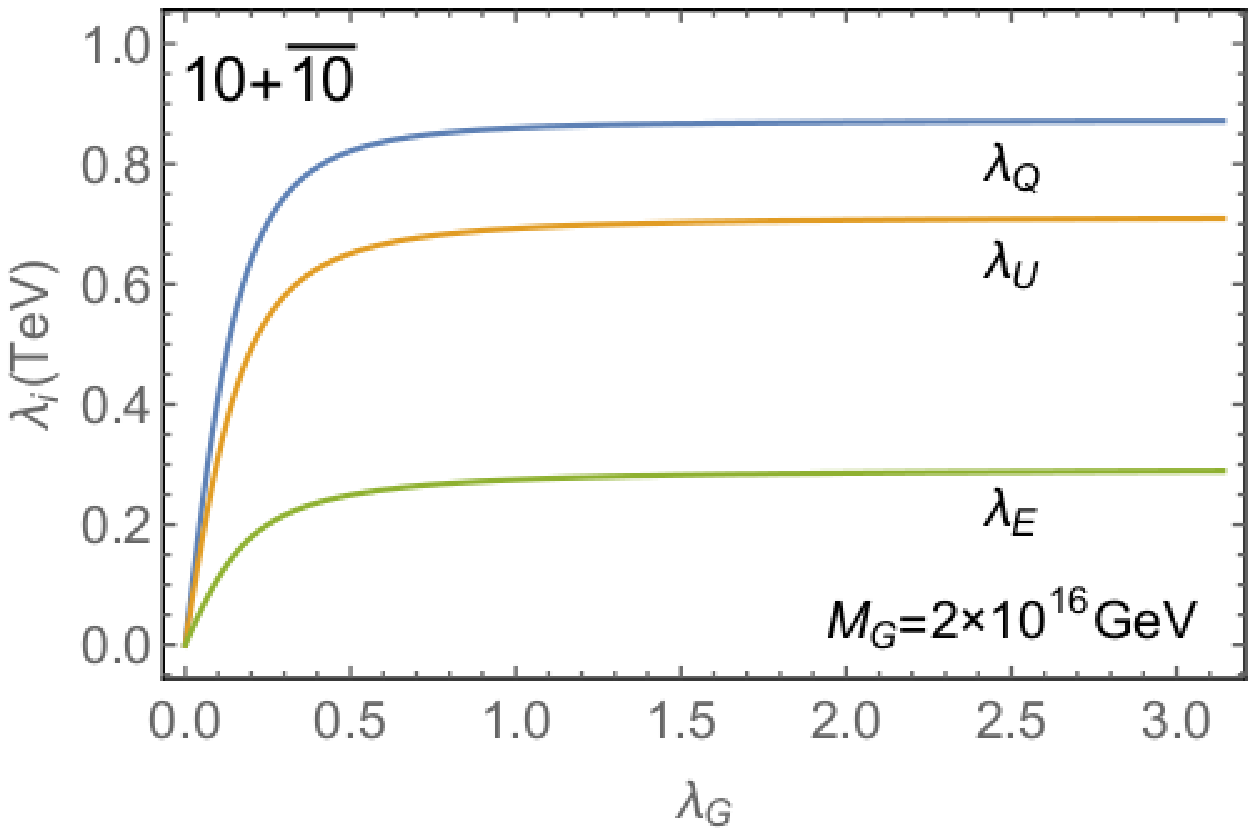}}
\end{center}
\caption{$\lambda_{D,L}$(TeV) for the $({\bf 5} + \overline{\bf 5})_1$ 
 theory (left panel) and $\lambda_{Q,U,E}$(TeV) in the ${\bf 10} + 
 \overline{\bf 10}$ theory (right panel) as a function of the unified 
 coupling $\lambda_{\rm G}$, computed from one-loop renormalization 
 group running.}
\label{fig:lambda}
\end{figure}
\begin{table}[t]
\begin{center}
\begin{tabular}{|l|l||c|c|c|c|c|}
\hline
    & & $D$ & $L$ & $Q$ & $U$ & $E$ \\ \hline\hline
  \multirow{2}{*}{$({\bf 5} + \overline{\bf 5})_1$} & $\lambda_i$ 
    & 0.96 & 0.63 & \multirow{2}{*}{---} & \multirow{2}{*}{---} 
    & \multirow{2}{*}{---} \\
    & $\mu_i/\mu_L$ & 1.5 & 1 & & & \\ \hline
  \multirow{2}{*}{$({\bf 5} + \overline{\bf 5})_2$} & $\lambda_i$ 
    & 0.77 & 0.46 & \multirow{2}{*}{---} & \multirow{2}{*}{---} 
    & \multirow{2}{*}{---} \\
    & $\mu_i/\mu_L$ & 1.7 & 1 & & & \\ \hline
  \multirow{2}{*}{$({\bf 5} + \overline{\bf 5})_3$} & $\lambda_i$ 
    & 0.70 & 0.36 & \multirow{2}{*}{---} & \multirow{2}{*}{---} 
    & \multirow{2}{*}{---} \\
    & $\mu_i/\mu_L$ & 1.9 & 1 & & & \\ \hline
  \multirow{2}{*}{$({\bf 5} + \overline{\bf 5})_4$} & $\lambda_i$ 
    & 0.67 & 0.27 & \multirow{2}{*}{---} & \multirow{2}{*}{---} 
    & \multirow{2}{*}{---} \\
    & $\mu_i/\mu_L$ & 2.5 & 1 & & & \\ \hline
  \multirow{2}{*}{${\bf 10} + \overline{\bf 10}$} & $\lambda_i$ 
    & \multirow{2}{*}{---} & \multirow{2}{*}{---} & 0.87 & 0.71 & 0.29 \\
    &  $\mu_i/\mu_E$ &  &  & 3.0 & 2.5 & 1 \\ \hline
  \multirow{2}{*}{${\bf 15} + \overline{\bf 15}$} & $\lambda_i$ 
    & 0.61 & 0.24  & 0.84 & 0.65 & 0.19 \\
    & $\mu_i/\mu_E$ & 3.2 & 1.3 & 4.5 & 3.4 & 1 \\ \hline
\end{tabular}
\caption{Predictions for $\lambda_i$(TeV) and physical mass ratios 
 $\mu_i/ \mu_{L,E}$ at one loop level.  The mass ratios assume a common 
 value for $\mu_i$ at $M_{\rm G}$.}
\label{table:couppred}
\end{center}
\end{table}
The production of $S(750)$ at LHC occurs via gluon fusion induced by 
virtual vector quarks, and the decay to photons is via loops of vector 
quarks and leptons.  An analysis of Run~1 and Run~2 data from both ATLAS 
and CMS leads to $\sigma B_{\gamma\gamma} \simeq (6 \pm 2)~{\rm fb}$ 
at $\sqrt{s} = 13~{\rm TeV}$~\cite{Buttazzo:2015txu}.  This requires 
couplings $\lambda_i$ at the TeV scale of order unity, and a key question 
is whether this is consistent with couplings not hitting a Landau pole up 
to $M_{\rm G}$.  In Figure~\ref{fig:lambda}, we plot $\lambda_{D,L}$(TeV) 
for the $({\bf 5} + \overline{\bf 5})_1$ theory (left panel) and 
$\lambda_{Q,U,E}$(TeV) in the ${\bf 10} + \overline{\bf 10}$ theory 
(right panel) as a function of the unified coupling $\lambda_{\rm G}$ 
as computed from one-loop renormalization group running.  (The effects 
from two loops on the low energy values of these couplings are small, 
$\lesssim 2\mbox{-}3\%$ in these theories and $\lesssim \mbox{a few }\%$ 
for $({\bf 5} + \overline{\bf 5})_4$ and ${\bf 15} + \overline{\bf 15}$.) 
A striking feature is that as $\lambda_{\rm G}$ becomes larger 
than unity so the low energy couplings lose their sensitivity 
to the high scale value.  Thus in most of the region where 
$\sigma B_{\gamma\gamma}$ is large enough to explain the diphoton 
data, our results are independent of $\lambda_{\rm G}$ and, 
indeed, independent of whether the couplings unify.  The predicted 
values of the TeV scale couplings in the various theories are given 
in Table~\ref{table:couppred}.  The splittings between the TeV scale 
values of $\lambda_i$ arise from running of the standard model gauge 
couplings and are calculable; QCD running ensures that the couplings 
to vector quarks are larger than to vector leptons.

\begin{figure}[t]
\begin{center}
  \includegraphics[scale=0.85]{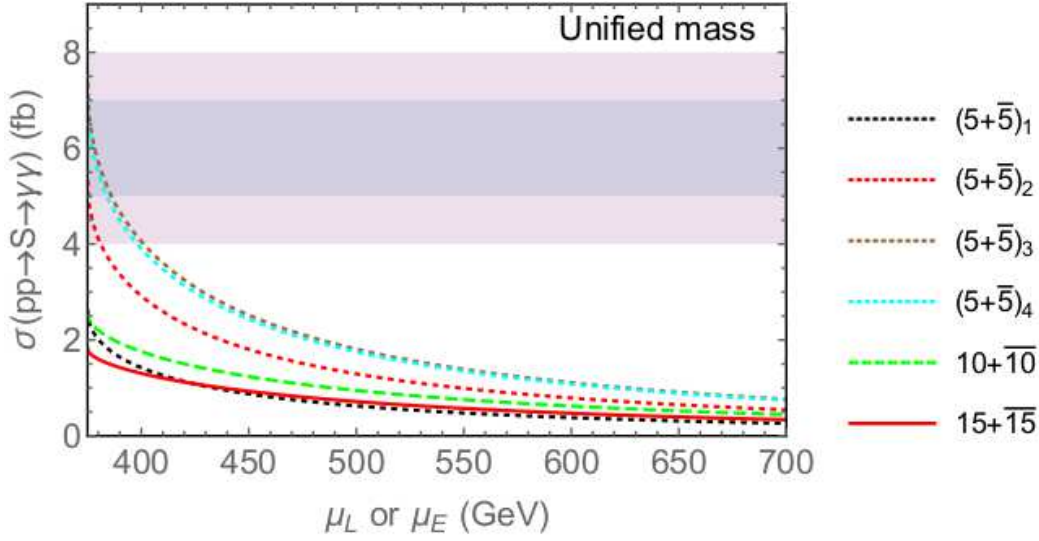}
\end{center}
\caption{Theories with unified mass relations: Prediction for 
 $\sigma B_{\gamma \gamma}$ at $\sqrt{s} = 13~{\rm TeV}$ as a function 
 of the lightest vector lepton mass.}
\label{fig:sigmaB-unif}
\end{figure}
For the case that $\mu_i$ unify at $M_{\rm G}$, the overall scale of 
$\mu_i$(TeV) depends on the unified mass, $\mu_{\rm G}$, while the 
splitting between $\mu_{D,L}$ or between $\mu_{Q,U,E}$ is again computed 
from gauge running.  Relative values for the physical masses $\mu_i$ 
are shown in Table~\ref{table:couppred} for each of our theories, 
assuming that they all take a common value at $M_{\rm G}$.  With 
these results, we compute $\sigma B_{\gamma \gamma}$ in terms of 
just one parameter, which we take to be the lightest vector lepton 
mass, as shown in Figure~\ref{fig:sigmaB-unif}.  In the numerical 
plots of this paper, we include contributions from both the scalar 
and pseudoscalar in $S$.  If the resonance is produced solely by the 
scalar (pseudoscalar) then $\sigma B_{\gamma \gamma}$ is reduced, 
by a factor of $4/13$ ($9/13$) in the heavy vector limit.  Also we 
determine the QCD $K$-factor (i.e.\ normalize the production cross 
section of scalar particles) so that it reproduces the production cross 
section of a standard~model-like Higgs boson of mass $750~{\rm GeV}$ 
at $\sqrt{s}=14~{\rm TeV}$~\cite{LHC-Higgs}.  Including this factor, 
and both scalar and pseudoscalar contributions, gives a local analytic 
approximation
\begin{equation}
  \sigma B_{\gamma \gamma} 
  \sim 1~{\rm fb} \left( \sum_i N_i \lambda_i Q_i^2\, 
    \frac{500~{\rm GeV}}{\mu_i} \right)^2,
\label{eq:sigmaapprox}
\end{equation}
where $\Phi_i$ contains $N_i$ states of electric charge $Q_i$. 
This local approximation is valid for $\mu_i \gtrsim 500~{\rm GeV}$; 
for lighter masses the dependence on $\mu_i$ is steeper, as seen 
in Figure~\ref{fig:sigmaB-unif}.

In the $({\bf 5} + \overline{\bf 5})_1$, ${\bf 10} + \overline{\bf 10}$, 
and ${\bf 15} + \overline{\bf 15}$ theories, the prediction is far 
too small to explain the observed value of $\sigma B_{\gamma \gamma}$, 
indicated by the horizontal shaded bands.  From Eq.~(\ref{eq:sigmaapprox}), 
the contribution from $\Phi_i$ is proportional to $Q_i^4$, leading to 
the importance of charged leptons and up-type quarks compared to down-type 
quarks.  Thus, it is at first surprising to see that the ${\bf 10} + 
\overline{\bf 10}$ and ${\bf 15} + \overline{\bf 15}$ theories fail 
to account for the data.  However, other effects are also important:\ 
Table~\ref{table:couppred} shows that as more vector states are added 
so the predicted value of the couplings $\lambda_i$(TeV) are reduced. 
This results from the increasing renormalization of the $S$ field and 
decreases the signal.  Furthermore, the splittings $\mu_i/\mu_{L,E}$ 
increase due to the larger value of the unified gauge coupling.  Since 
$\mu_{L,E}$ must be larger than $\mu_S/2$, this increases the masses 
of the vector quarks, again decreasing the signal.  The $({\bf 5} 
+ \overline{\bf 5})_{2,3,4}$ theories can account for the diphoton 
resonance, but only if there are 2, 3 or 4 charged vector leptons lighter 
than about $400~{\rm GeV}$.  In this case, the $D$ vector quarks are 
near $1~{\rm TeV}$.  This avoids the bounds on vector quarks from Run~1 
even if they decay into the third generation quarks~\cite{Aad:2015kqa}. 
The bounds are weaker if they decay mainly into the first two 
generation quarks~\cite{Aad:2015tba}.

\begin{figure}[t]
\begin{center}
  \subfigure{\includegraphics[clip,width=.48\textwidth]{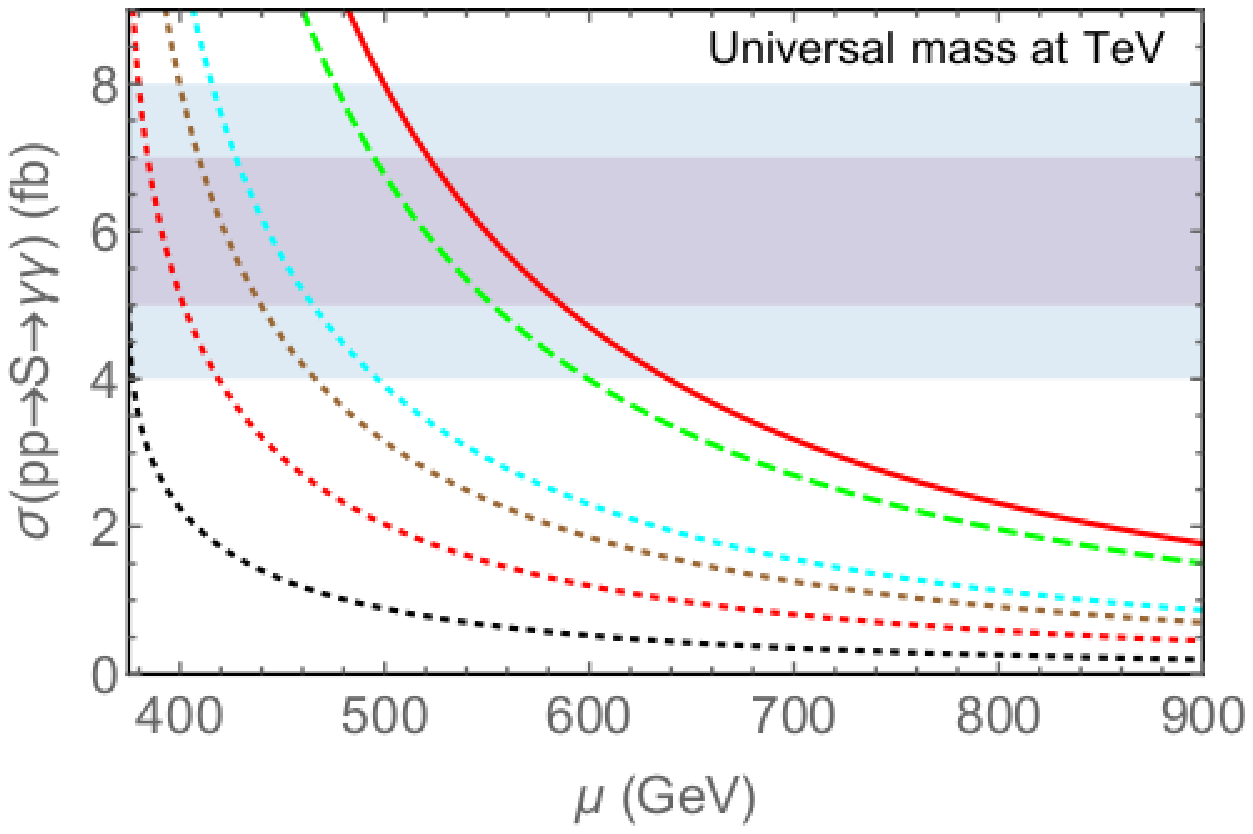}}
  \quad
  \subfigure{\includegraphics[clip,width=.48\textwidth]{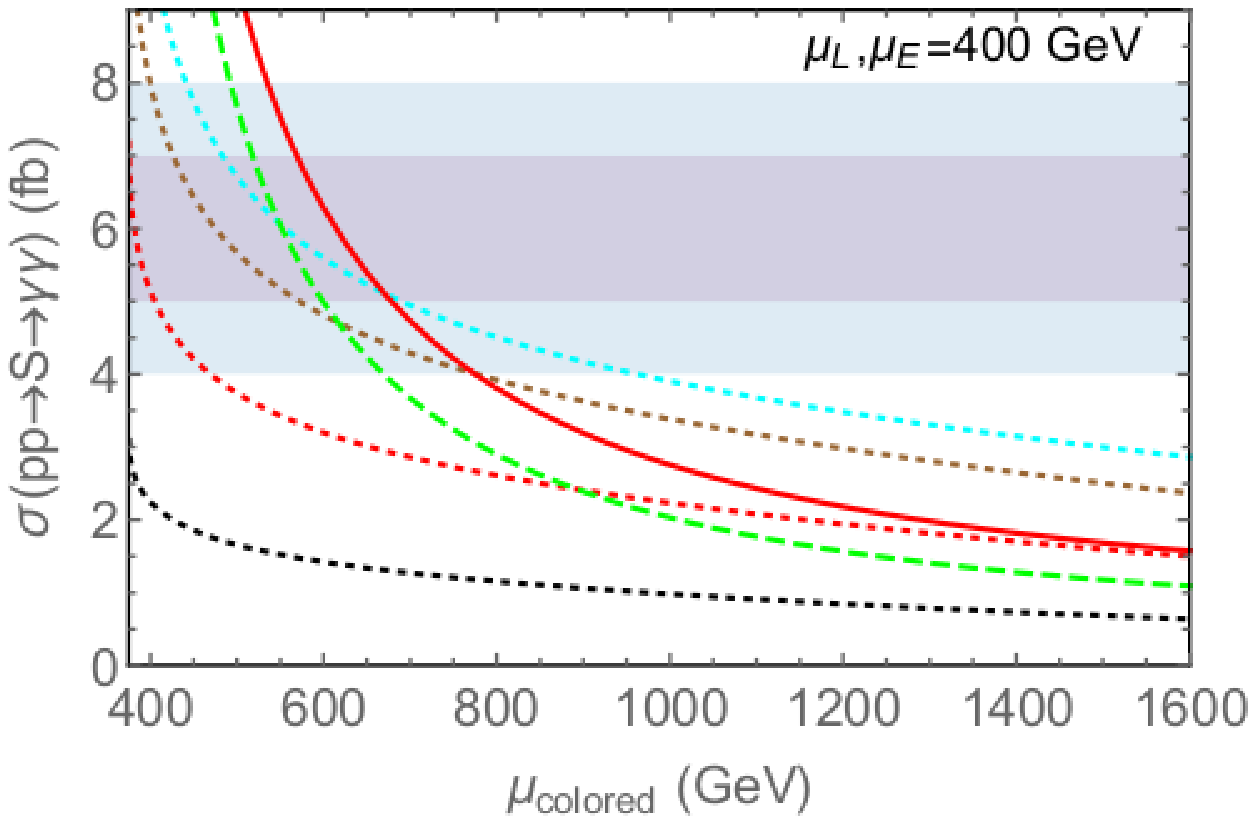}}
\end{center}
\caption{Theories without unified mass relations: Prediction for 
 $\sigma B_{\gamma \gamma}$ at $\sqrt{s} = 13~{\rm TeV}$ as a function 
 of (i) the degenerate mass of all vector quarks and leptons (left 
 panel), (ii) the degenerate mass of the vector quarks, with vector 
 lepton masses fixed at $400~{\rm GeV}$ (right panel).  The definitions 
 of colored lines are the same as in Figure~\ref{fig:sigmaB-unif}.}
\label{fig:sigmaB-var}
\end{figure}
Unified boundary conditions on $\lambda_i$ play no role under the common 
phase assumption, since we need large couplings leading to insensitivity 
to the ultraviolet.  However, unified boundary conditions on $\mu_i$ 
are extremely constraining:\ most regions of parameter space are far 
from explaining the $750~{\rm GeV}$ diphotons, and only a narrow range 
of vector lepton masses between $375~{\rm GeV}$ and $400~{\rm GeV}$ in 
the $({\bf 5} + \overline{\bf 5})_{2,3,4}$ theories account for the data. 
Next we explore the case without unified boundary conditions on $\mu_i$. 
This readily happens in theories where boundary conditions in extra 
dimensions break the unified symmetry~\cite{Hall:2001pg}, but it can 
also occur in 4~dimensions if these masses pick up unified symmetry 
breaking effects at an $O(1)$ level.%
\footnote{If the masses $\mu_i$ are generated mainly by terms involving 
 a condensation(s) of a unified symmetry breaking field(s), then mass 
 splittings among fields in a single unified group multiplet can be 
 $O(1)$.  This can happen, for example, if the unified symmetry breaking 
 field is charged under some symmetry.}
Predictions for $\sigma B_{\gamma\gamma}$ are shown in 
Figure~\ref{fig:sigmaB-var} for all the theories for two sample spectra. 
In the left panel, all vector quarks and leptons are taken degenerate, 
opening up a large region for the ${\bf 10} + \overline{\bf 10}$ and 
${\bf 15} + \overline{\bf 15}$ theories with masses up to $600~{\rm GeV}$ 
or more.  Vector quarks of these masses are not excluded if they decay 
mainly into the first two generation quarks~\cite{Aad:2015tba}.  From  
Eq.~(\ref{eq:sigmaapprox}), we see that both the charged leptons and 
charge 2/3 quarks are particularly important, and it is clear that 
degeneracy of the states is not required but is a simple choice for 
illustration.  Even without unified boundary conditions on $\mu_i$, 
QCD running tends to make the vector quarks heavier than the vector 
leptons.  Hence in the right panel, we hold the vector lepton masses 
fixed at $400~{\rm GeV}$ and allow the vector quark masses to vary 
to higher values.  This allows even heavier vector quarks, up to mass 
$700~{\rm GeV}$ for ${\bf 10} + \overline{\bf 10}$, $800~{\rm GeV}$ 
for ${\bf 15} + \overline{\bf 15}$, and about a TeV for $({\bf 5} 
+ \overline{\bf 5})_{3,4}$.  Relaxing the condition of unified $\mu_i$ 
greatly expands the possibilities for explaining the $750~{\rm GeV}$ 
diphoton data, and leads to the exciting expectation of many 
vector quarks and leptons in reach of LHC in the region of 
$400~{\rm GeV}$--$1~{\rm TeV}$.

In generic gravity mediation, we expect that the soft parameter $b_S$ 
is of order the square of the supersymmetry breaking scale, so that the 
scalar and pseudoscalar components of $S$ are well split.  In this case, 
all the results given above for $\sigma B_{\gamma\gamma}$ should be 
multiplied by about $4/13$ ($9/13$) for the $750~{\rm GeV}$ state 
identified as the scalar (pseudoscalar).  However, if the mediation 
mechanism leads to a suppression of $b$ parameters, the splitting may 
be smaller; in this case there is an irreducible one-loop contribution 
arising from virtual $\Phi_i$, leading to a mass splitting between the 
scalar and pseudoscalar of
\begin{align}
  \varDelta m_{\rm irred}
  &\simeq \frac{1}{4\pi^2} \frac{1}{2 m_S} 
    \sum_i N_i \lambda_i^2 \mu_i^2\, \ln \frac{\mu_i^2+m_i^2}{\mu_i^2}
\nonumber\\
  &\simeq 15\mbox{--}30~{\rm GeV} 
    \left( \frac{\mu_i}{500~{\rm GeV}} \right)^2 
    \ln \frac{\mu_i^2+m_i^2 }{\mu_i^2},
\label{eq:PSsplitting}
\end{align}
where $m_i^2$ is the soft supersymmetry breaking mass parameter for 
$\Phi_i$, and in the last expression we have assumed a common value for 
$\mu_i$ and $m_i^2$, and used $\lambda_i$ from Table~\ref{table:couppred}.  
The pre-factor range of $15\mbox{--}30~{\rm GeV}$ reflects the range 
obtained in the various theories.  This could easily lead to a mass 
splitting of 10s of GeV, so that more data will reveal two nearby 
resonances, with one having a signal about $9/4$ of the other, for 
sufficiently large $\mu_i$.  This could account for the current mild 
preference for a width of about $40~{\rm GeV}$.  Finally, as $m_i^2$ 
drops below $\mu_i^2$ a supersymmetric cancellation suppresses $\varDelta 
m_{\rm irred}$, so that a narrow resonance emerges with width below 
the experimental resolution.  In this case, and in the case of two 
nearby resonances, the signal to be compared with current data is the 
sum of scalar and pseudoscalar contributions, as presented in our plots. 

For vector squarks/sleptons heavier than $1~{\rm TeV}$ the corrections 
to the diphoton rate are very small; below a TeV they give a subdominant 
contribution to the diphoton signal, slightly weakening our bounds 
on the upper limit on the vector quark/lepton masses, for example 
by about $10\%$.  We note that if $\lambda_i$ have different phases, 
the various contributions in Eq.~(\ref{eq:PSsplitting}) may partially 
cancel.  Furthermore, with a phase in $b_S$ the scalar and pseudoscalar 
from $S$ mix, affecting the diphoton signal strength.

Finally, we mention the possibility that vector matter parity is unbroken 
leading to stability of the lightest vector fermion, which could be 
dark matter.  This is of particular interest for the case of $R$-parity 
violation where the lightest supersymmetric particle cannot be dark 
matter.  With vector matter parity, flavor changing masses via the Higgs 
expectation value allow heavy vector quarks to cascade to lighter ones, 
and similarly for vector leptons; however, this still leads to stability 
of the lightest vector quark and lepton.  However, $R$-parity violating 
operators such as $\Phi_Q \Phi_L D$ allow vector quarks to decay to 
vector leptons, so only the lightest vector lepton is stable.  In theories 
containing a neutral vector lepton $\Phi_N$ as well as the doublet 
$\Phi_L$, the Yukawa coupling $\Phi_L \Phi_N H$ mixes the doublet and 
singlet vector leptons leading to the well known simple possibility of 
``Singlet-Doublet'' dark matter~\cite{Mahbubani:2005pt,Elor:2009jp}.

Summarizing, the diphoton resonance can be explained by the addition of 
vector quarks and leptons and a singlet to the MSSM.  Requiring perturbativity 
to unified scales implies that several such states are sufficiently light 
to yield multiple signals at the LHC.  Direct pair production yields 
signals from the decay of each to a quark/lepton together with a $W,Z$ 
or standard model Higgs boson.  In addition, signals of the $750~{\rm GeV}$ 
resonance will be visible in $gg$, $Z\gamma$, $ZZ$, and $WW$ modes as 
well as $\gamma\gamma$ (see the Appendix).  In cases where fewer than 
five $\mu_i$ are independent, the signal strength of these modes will 
be over-constrained, allowing us to test consistency of the theory. 
Furthermore, if the new vector matter is discovered and $\mu_i$ are 
measured, then we could determine (some of) $\lambda_i$ from the rates 
of these two gauge boson modes and see if they are consistent with 
the assumption of large $\lambda_i$ at the unification scale.  Such 
discoveries would provide a new window to high scales that could 
strengthen the case for supersymmetric unification and have important 
implications for unified theories.  Other predictions of unified theories 
would also be impacted, such as quark and lepton mass relations and 
proton decay.

\section*{Acknowledgments}
We thank N.~Yokozaki for bringing our attention to the issue 
of strong coupling in the $({\bf 5} + \overline{\bf 5})_4$ and 
${\bf 15} + \overline{\bf 15}$ theories.  This work was supported 
in part by the Director, Office of Science, Office of High Energy 
and Nuclear Physics, of the U.S.\ Department of Energy under Contract 
DE-AC02-05CH11231, by the National Science Foundation under grants 
PHY-1316783 and PHY-1521446, and by MEXT KAKENHI Grant Number 15H05895.

\appendix

\section{Other Decay Modes of the Singlet}

\begin{figure}[t]
\begin{center}
  \subfigure{\includegraphics[clip,width=.35\textwidth]{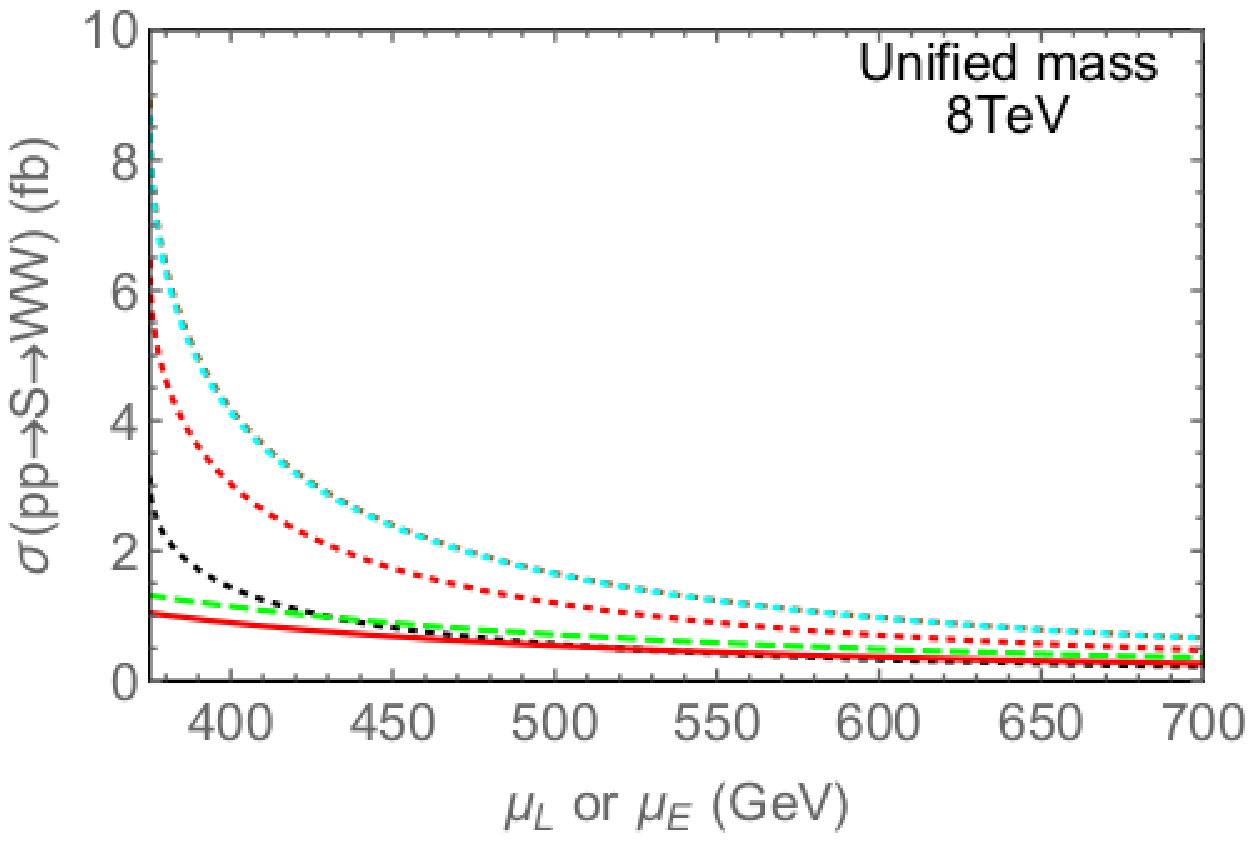}}
  \quad
  \subfigure{\includegraphics[clip,width=.35\textwidth]{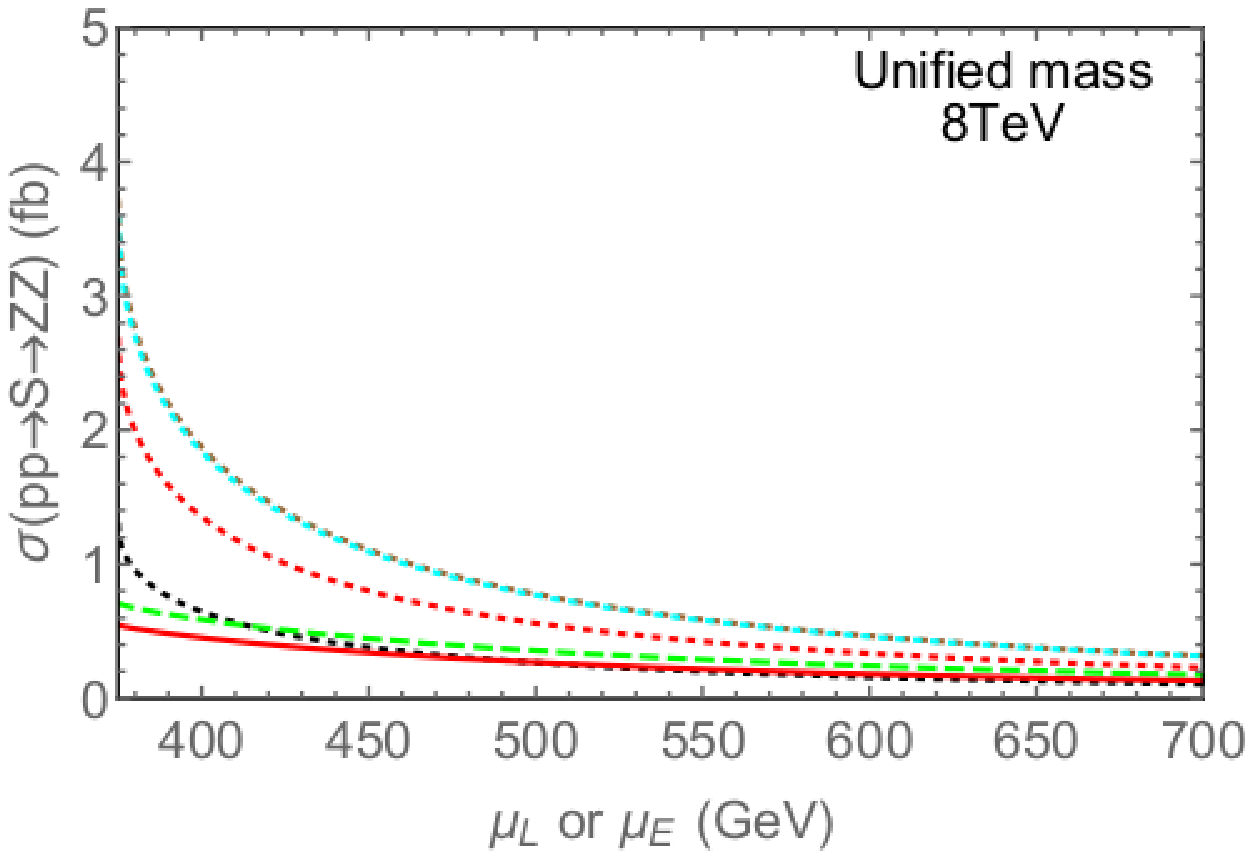}}
    \subfigure{\includegraphics[clip,width=.35\textwidth]{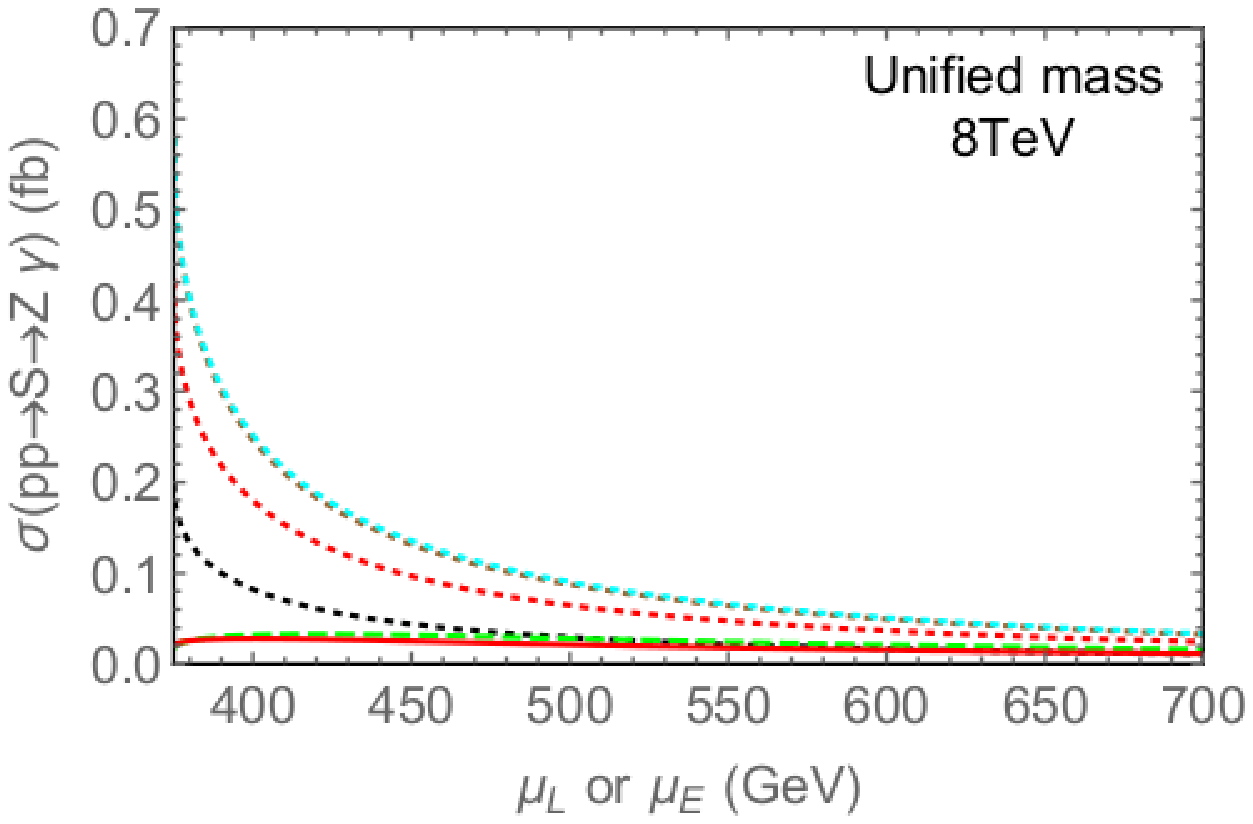}}
  \quad
  \subfigure{\includegraphics[clip,width=.35\textwidth]{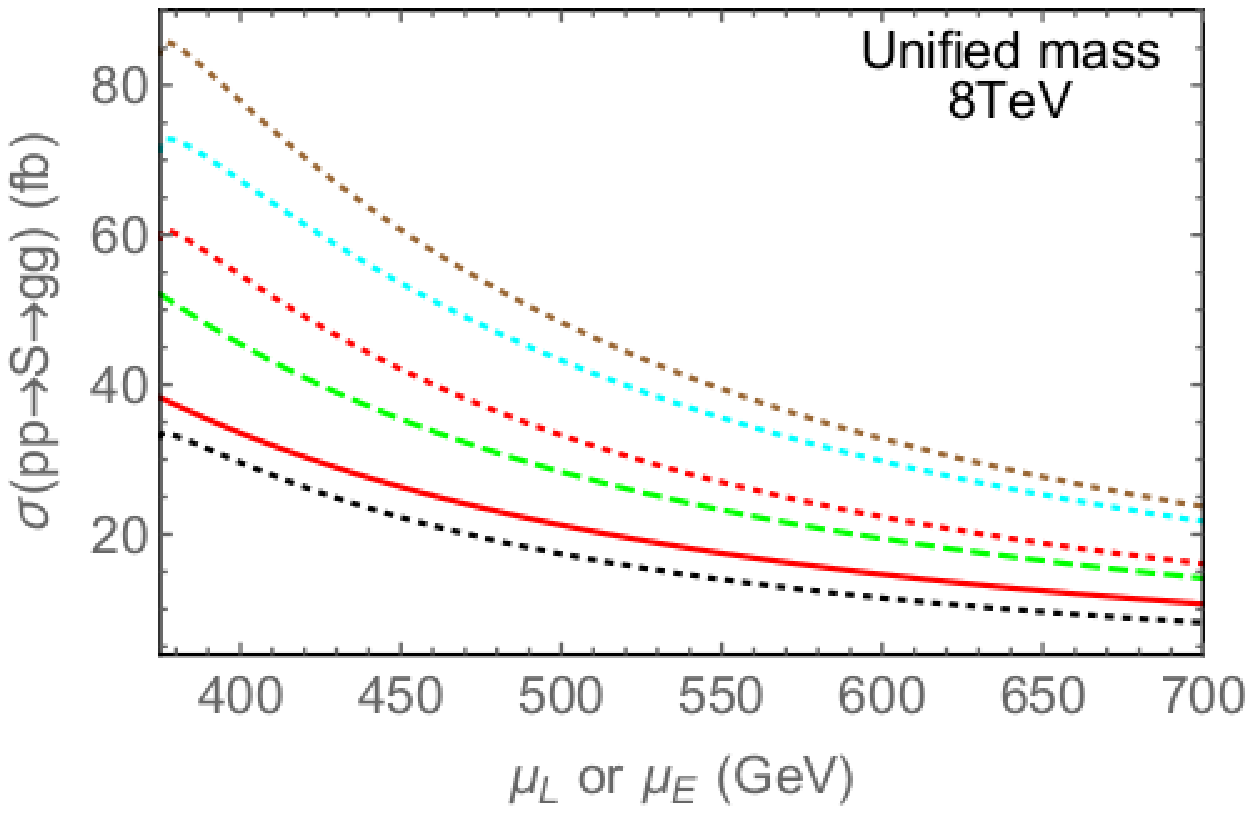}}
\end{center}
\caption{Theories with unified mass relations:\ Predictions for 
 $\sigma(pp \rightarrow S)$ times branching ratios into $WW$, $ZZ$, 
 $Z\gamma$, and $gg$ at $\sqrt{s} = 8~{\rm TeV}$ as a function of 
 the lightest vector lepton masses.  The definitions of colored lines 
 are the same as those in Figure~\ref{fig:sigmaB-unif}.}
\label{fig:other_unify}
\end{figure}

\begin{figure}[t]
\begin{center}
  \subfigure{\includegraphics[clip,width=.35\textwidth]{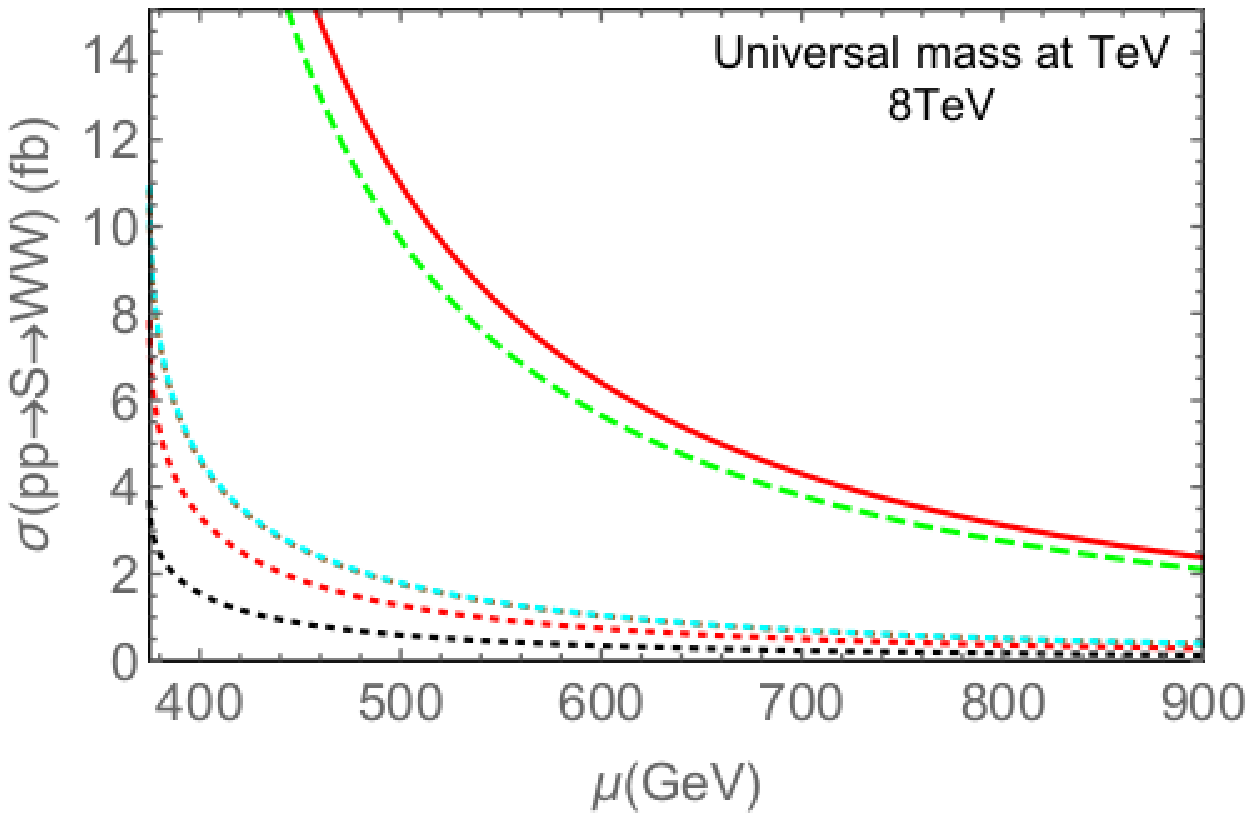}}
  \quad
  \subfigure{\includegraphics[clip,width=.35\textwidth]{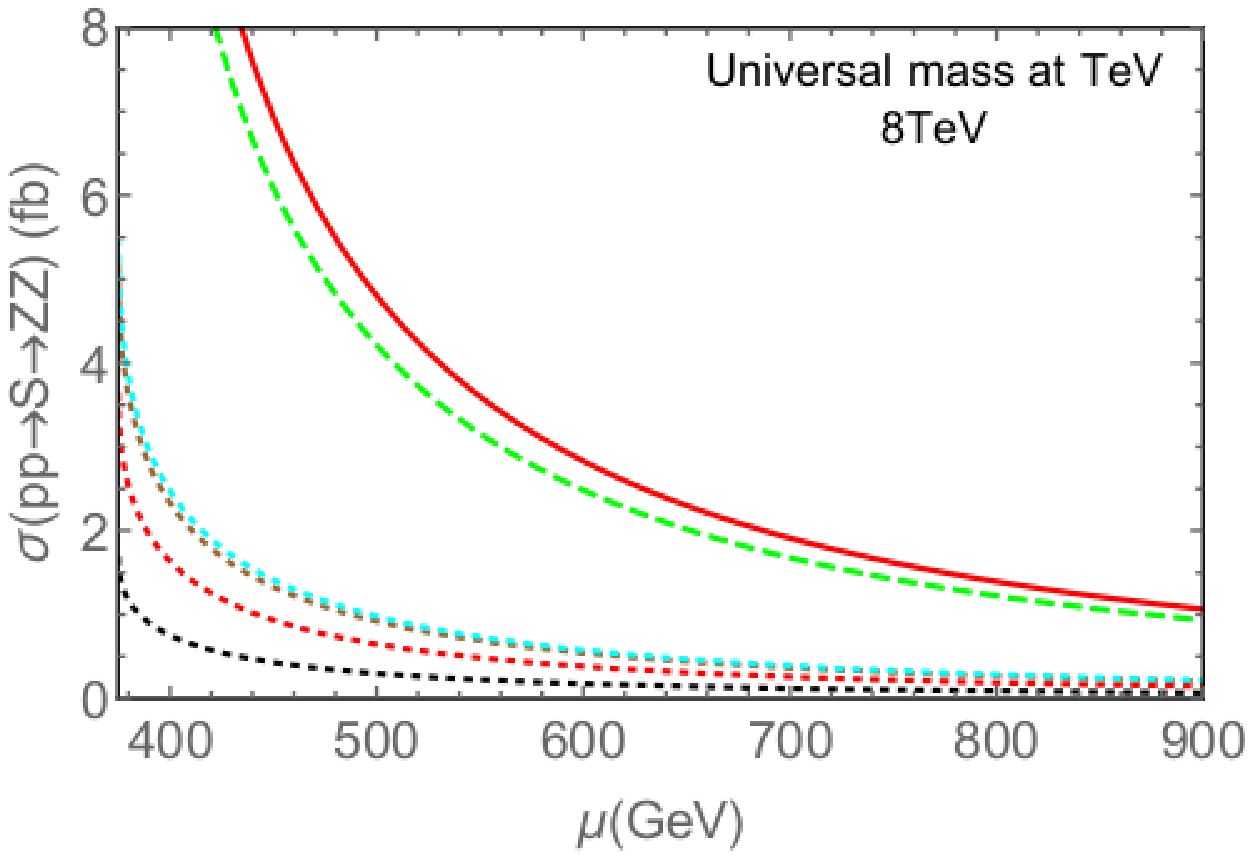}}
    \subfigure{\includegraphics[clip,width=.35\textwidth]{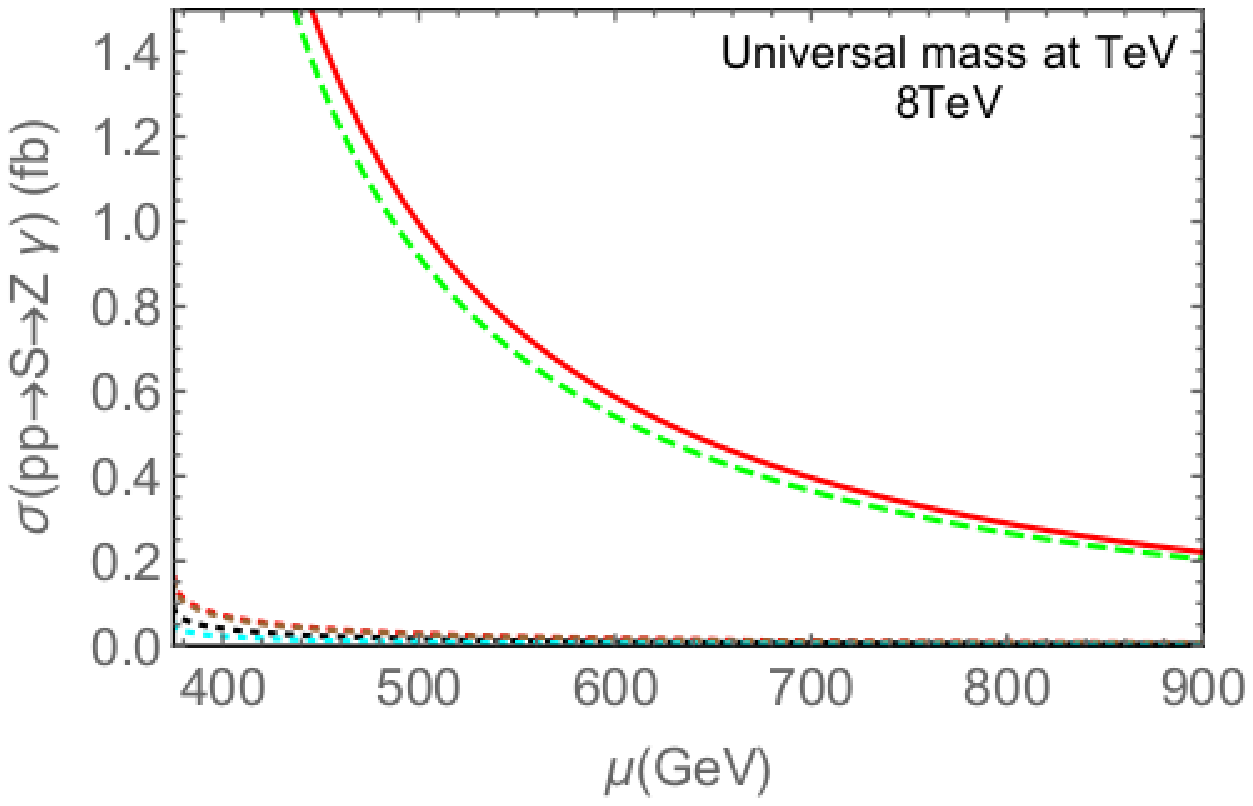}}
  \quad
  \subfigure{\includegraphics[clip,width=.35\textwidth]{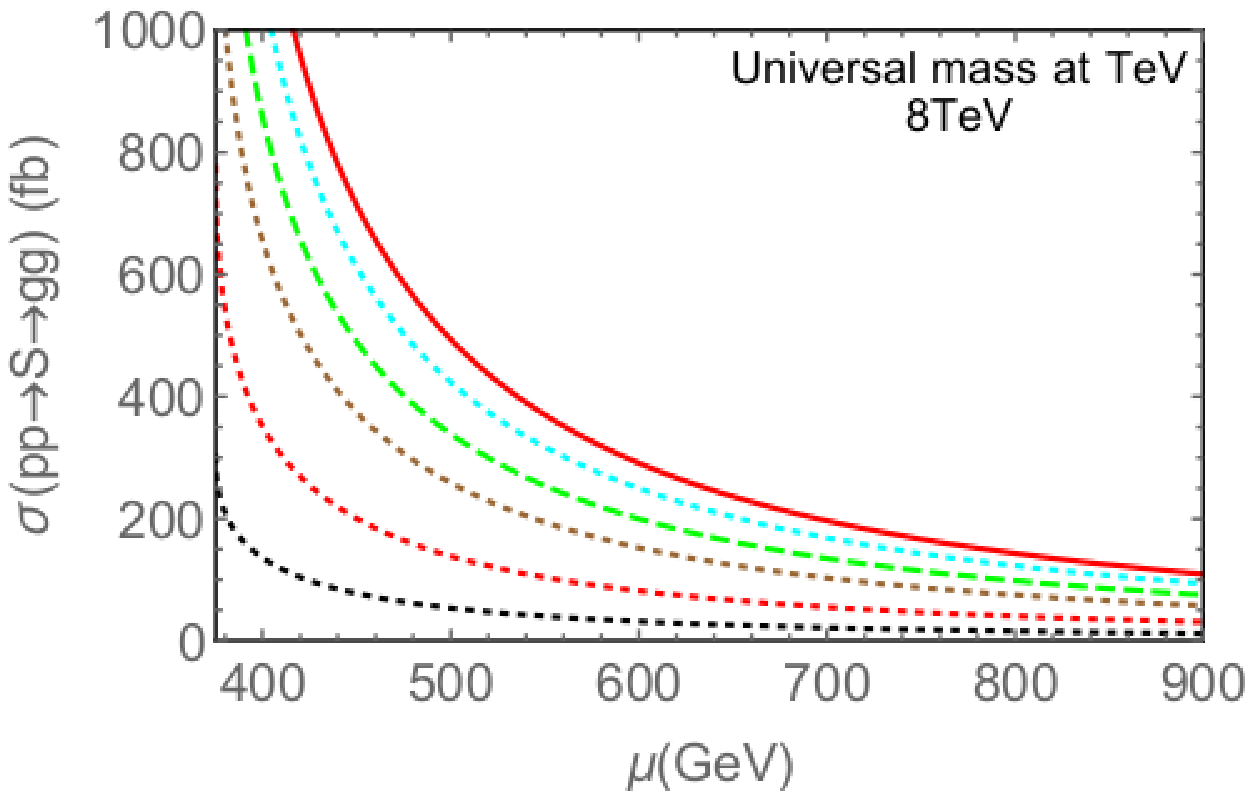}}
\end{center}
\caption{Theories without unified mass relations:\ Predictions for 
 $\sigma(pp \rightarrow S)$ times branching ratios into $WW$, $ZZ$, 
 $Z\gamma$, and $gg$ at $\sqrt{s} = 8~{\rm TeV}$ as a function of the 
 degenerate mass of all the vector quarks and leptons.  The definitions 
 of colored lines are the same as those in Figure~\ref{fig:sigmaB-unif}.}
\label{fig:other_univ}
\end{figure}

\begin{figure}[t]
\begin{center}
  \subfigure{\includegraphics[clip,width=.35\textwidth]{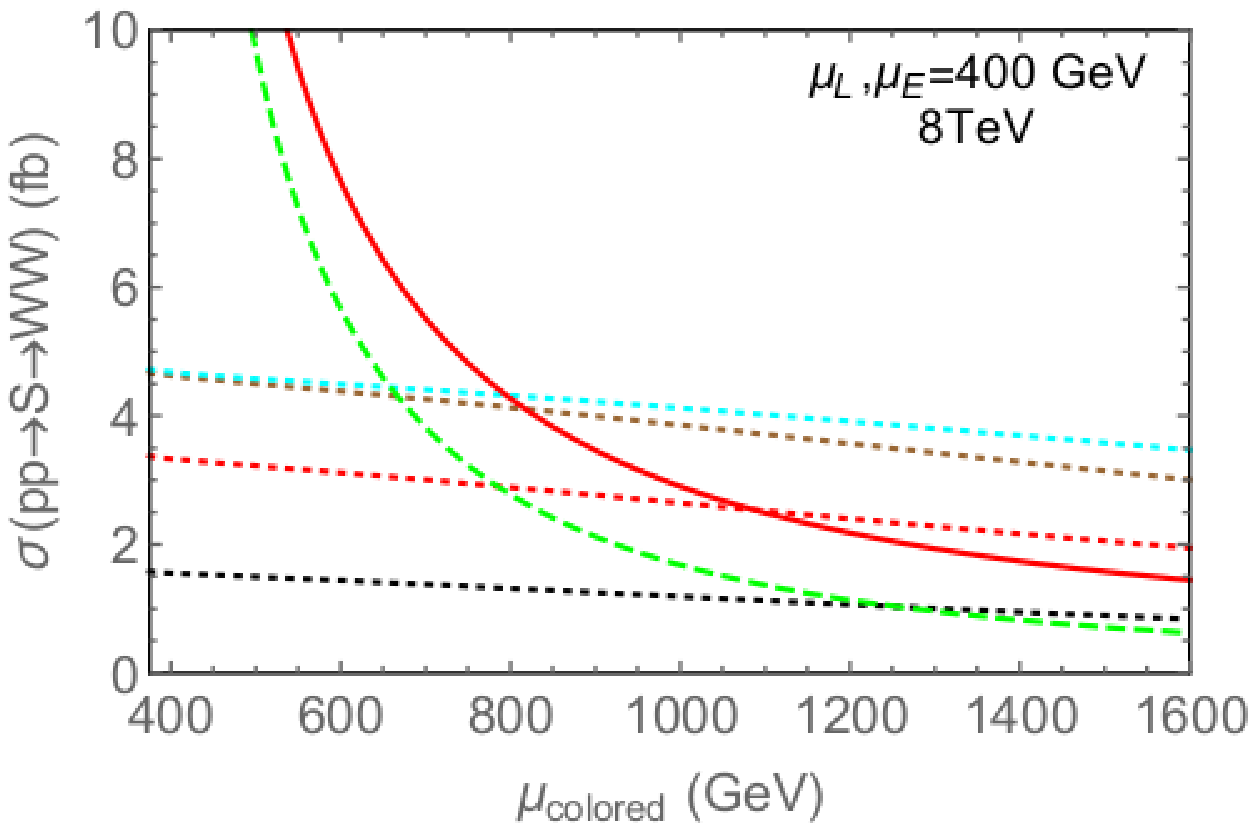}}
  \quad
  \subfigure{\includegraphics[clip,width=.35\textwidth]{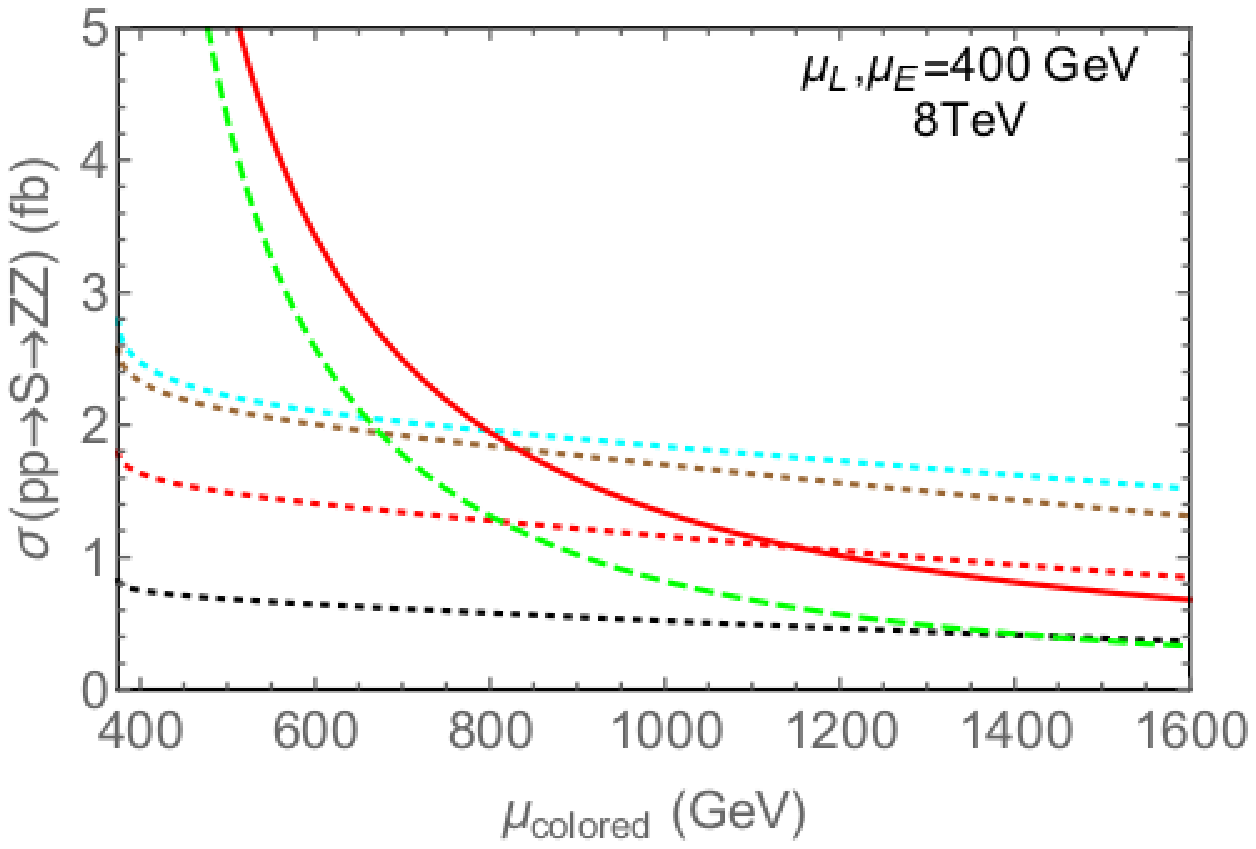}}
    \subfigure{\includegraphics[clip,width=.35\textwidth]{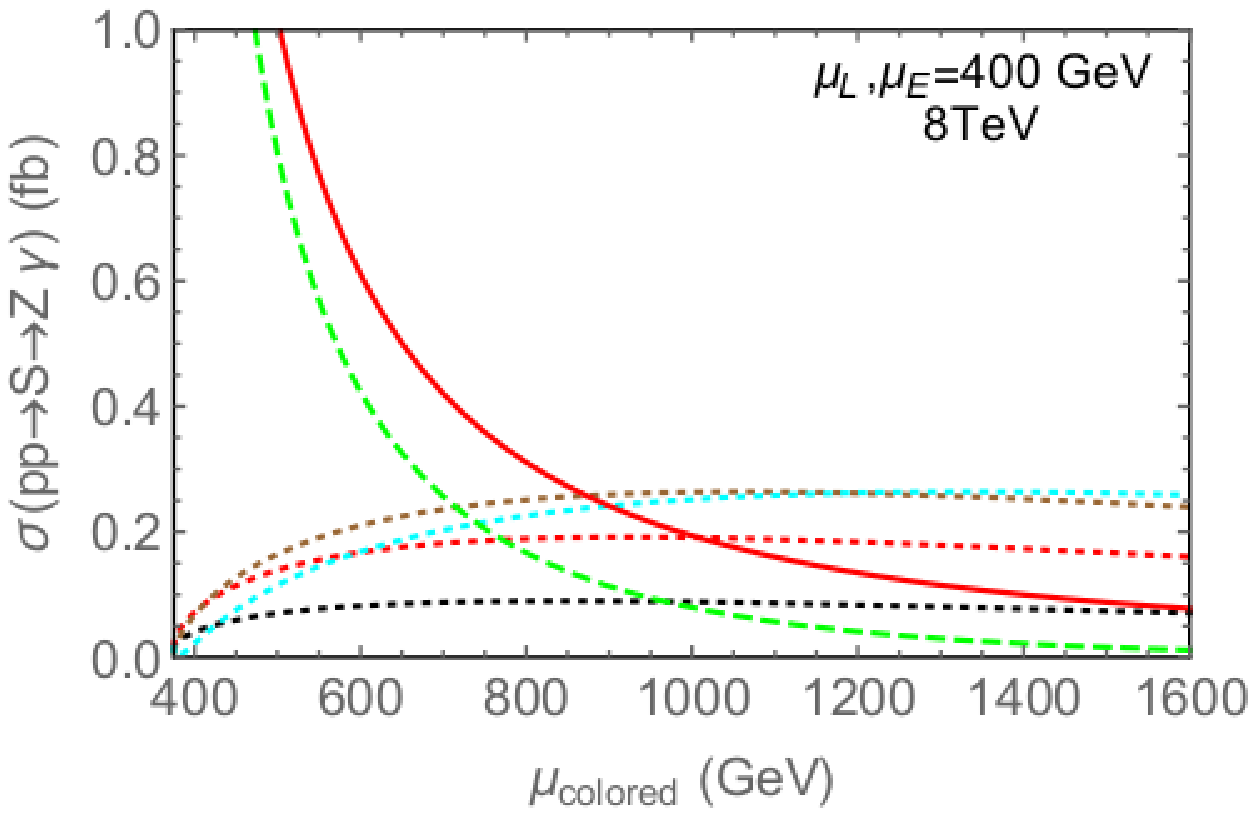}}
  \quad
  \subfigure{\includegraphics[clip,width=.35\textwidth]{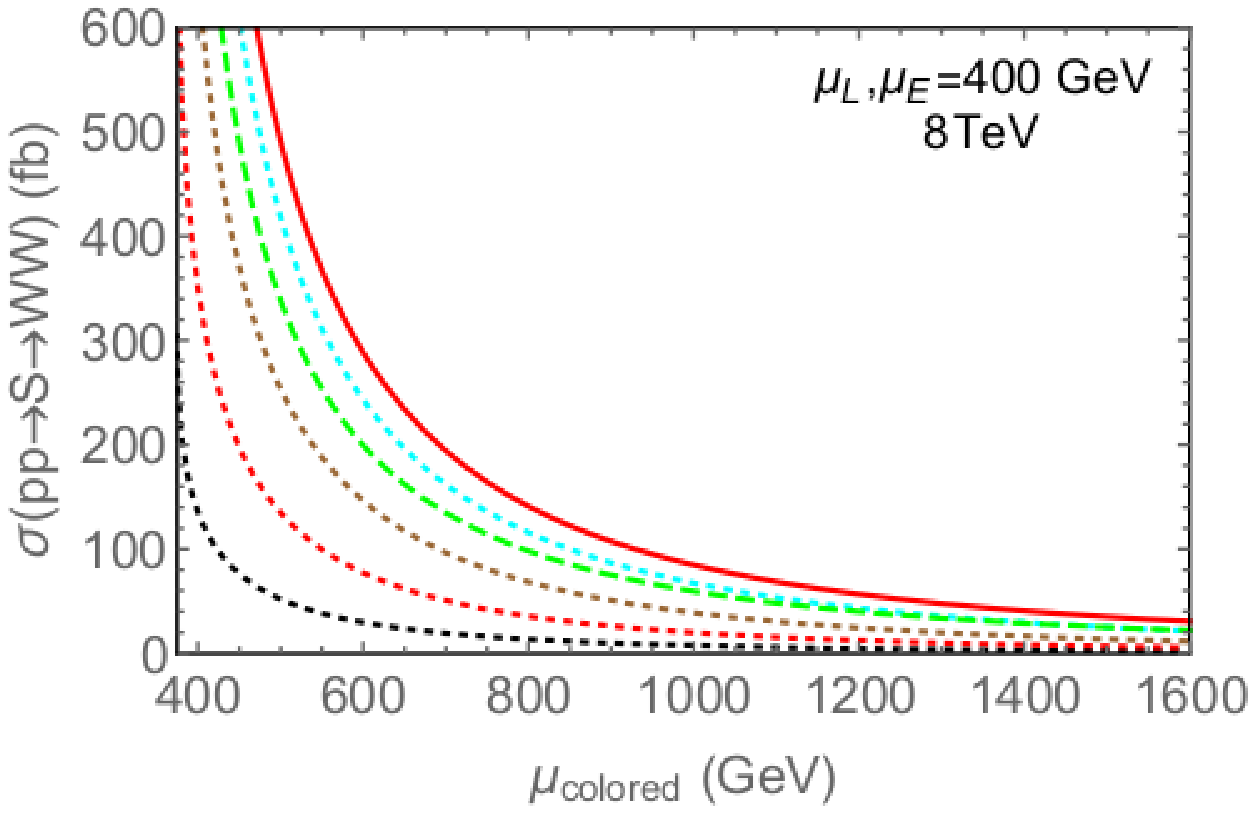}}
\end{center}
\caption{Theories without unified mass relations:\ Predictions for 
 $\sigma(pp \rightarrow S)$ times branching ratios into $WW$, $ZZ$, 
 $Z\gamma$, and $gg$ at $\sqrt{s} = 8~{\rm TeV}$ as a function of the 
 degenerate mass of the vector quarks, with the vector lepton masses 
 fixed at $400~{\rm GeV}$.  The definitions of colored lines are the 
 same as those in Figure~\ref{fig:sigmaB-unif}.}
\label{fig:other_opt}
\end{figure}

In this appendix, we show predictions for the production cross sections 
of $WW$, $ZZ$, $Z\gamma$, and $gg$ through the decay of the singlet $S$. 
Figures~\ref{fig:other_unify}, \ref{fig:other_univ}, and \ref{fig:other_opt}  show predictions of theories at the 8 TeV LHC,
respectively for the cases
with unified mass relations,
degenerate masses for all the vectorlike quarks and leptons at the TeV scale,
and degenerate masses for the vector quarks with lepton masses fixed at $400~{\rm GeV}$.
In the parameter regions consistent with 
the observed diphoton rate, the constraints from those decay 
modes~\cite{Aad:2015agg,Aad:2015kna,Aad:2014fha,dijet} are all 
satisfied.  The production cross sections at the $13~{\rm TeV}$ 
LHC is $4.8$ times those at the $8~{\rm TeV}$ LHC.

\end{document}